\documentclass[10pt,letterpaper]{article}
\usepackage[top=0.85in,left=2.75in,footskip=0.75in]{geometry}

\usepackage{amsmath,amssymb}

\usepackage{changepage}

\usepackage{textcomp,marvosym}

\usepackage{cite}

\usepackage{nameref,hyperref}

\usepackage[right]{lineno}

\usepackage[nopatch=eqnum]{microtype}
\DisableLigatures[f]{encoding = *, family = * }

\usepackage[table]{xcolor}

\usepackage{array}

\newcolumntype{+}{!{\vrule width 2pt}}

\newlength\savedwidth

\raggedright
\usepackage[aboveskip=1pt,labelfont=bf,labelsep=period,justification=raggedright,singlelinecheck=off]{caption}

\makeatletter
\renewcommand{\@biblabel}[1]{\quad#1.}
\makeatother

\usepackage{lastpage,fancyhdr,graphicx}
\usepackage{epstopdf}
\fancyheadoffset[L]{2.25in}
\fancyfootoffset[L]{2.25in}
\def\bbbr{{\mathbb R}}

\def\norm{\scriptsize\mbox{norm}}
\def\simple{\scriptsize\mbox{simple}}

\begin{document}
\vspace*{0.2in}

\begin{flushleft}
{\Large
\textbf{Orientation selectivity properties for integrated affine quasi quadrature models of complex cells} 
}
\newline
\\
Tony Lindeberg\textsuperscript{1}

\bigskip
\textbf{1} Computational Brain Science Lab,
        Division of Computational Science and Technology,
        KTH Royal Institute of Technology,
        SE-100 44 Stockholm, Sweden. 
\\

%
%





*tony@kth.se
\end{flushleft}

\section*{Abstract}

This paper presents an analysis of the orientation selectivity
properties of idealized models of complex cells in terms of affine
quasi quadrature measures, which combine the responses of idealized
models of simple cells in terms of affine Gaussian derivatives by
(i)~pointwise squaring, (ii)~summation of responses for different orders
of spatial derivation and (iii)~spatial integration.
Specifically, this paper explores the consequences of assuming that
the family of spatial receptive fields should be covariant under
spatial affine transformations, thereby implying that the receptive
fields ought to span a variability over the degree of elongation.
We investigate the theoretical properties of three main ways of
defining idealized models of complex cells and
compare the predictions from these models to neurophysiologically
obtained receptive field histograms over the resultant of biological
orientation selectivity curves. It is shown that the extended
modelling mechanisms lead to more uniform behaviour and a wider span
over the values of the resultant that are covered, compared
to an earlier presented idealized model of complex cells without
spatial integration.

More generally, we propose to, based on the presented results: (i)~include an explicit
variability over the degree of elongation of the receptive fields
in functional models
of complex cells, and that (ii)~the
suggested methodology with comparisons to biological
orientation selectivity curves and orientation selectivity
histograms
could be used as a new tool to evaluate other
computational models of complex cells in relation to biological
measurements.

\section*{Introduction}

To understand the functional properties of the visual system, it is
essential to aim at bridging the gap between computational models on one
side and neurophysiological measurements on the other side.
Specifically, whenever possible, it is desirable to construct 
theoretically principled models,
which could then lead to a deeper
understanding of visual processing modules and also generate
predictions for further biological experiments.
Regarding visual perception, it is in particular important to develop
a good understanding of the receptive fields%
\footnote{According to the pioneering work by Hubel and Wiesel
\cite{HubWie59-Phys,HubWie62-Phys,HubWie68-JPhys,HubWie05-book},
a visual receptive field is defined as the region in visual space
that can contribute to the response of a visual neuron.
In this work, we follow a functional extension of that approach, by
defining a visual receptive field as the computational function
that computes the response of a visual neuron to the visual stimuli
within the support region of the receptive field, that is over the
region in the visual space that can evoke a response of the neuron.}
in the early layers of
the visual hierarchy, which constitute the fundamental primitives
that the higher layers in the visual hierarchy are based upon.

One area where it
indeed seems to be possible to bridge the gap between neurophysiological
recordings and principled theory concerns the normative theory
for visual receptive fields in Lindeberg \cite{Lin21-Heliyon}. This theory
has been developed from principled assumptions regarding
symmetry properties of an idealized vision system, and
leads to a canonical family of linear receptive fields in
terms of spatial derivatives of affine Gaussian kernels.
Interestingly, the shapes of these idealized receptive fields
do rather well correspond to the qualitative shapes of
simple cells recorded by DeAngelis {\em et al.\/}\
\cite{DeAngOhzFre95-TINS,deAngAnz04-VisNeuroSci},
Conway and Livingstone \cite{ConLiv06-JNeurSci} and
Johnson {\em et al.\/}\ \cite{JohHawSha08-JNeuroSci};
see Fig.~12--18 in Lindeberg \cite{Lin21-Heliyon} for 
comparisons between biological receptive fields and idealized models.

One of the components in this normative theory for visual receptive
fields is the assumption that the family of receptive fields ought to
be covariant%
\footnote{The notion of a receptive field being covariant under
  geometric image transformations means that the receptive field model
  is well-behaved under the given class of geometric image
  transformation. If we let ${\cal G}$ denote the operator that
  computes a geometrically transformed image $f'$ from a given input
  image $f$ according to $f' = {\cal G} \, f$, then a receptive field
  represented by the operator ${\cal R}$ is said to be covariant under
  the corresponding class of geometric image transformations, if the
  result of applying the receptive field to the geometrically
  transformed image ${\cal R} \, {\cal G} f$ can be written as
  the result of applying a geometric transformation to the result of a
  related receptive field operator ${\cal R}'$ to the same input image
  according to ${\cal R} \, {\cal G} f = {\cal G} \, {\cal R}' \,
  f$. In this context, the related receptive field operator ${\cal R}'$ should
  either be a different member of the receptive field family defined
  by the operator ${\cal R}$, or constituting a sufficiently simple
  transformation of a receptive field defined by the operator ${\cal R}$.}
under spatial affine transformations, so as to enable more robust
processing of the image data as variations in the viewing conditions
imply variabilities in the image data caused by natural image
transformations
(see Lindeberg \cite{Lin23-FrontCompNeuroSci,Lin25-BICY,Lin25-JMIV}).
Specifically, if a visual observer views the same object
from different distances and viewing directions, then the local image
patterns will be deformed by the varying parameters of the perspective transformations,
which to first order can be approximated by local affine
transformations. In the area of computer vision, it has specifically
been shown that the property of covariance under spatial affine
transformations, referred to as affine covariance, enables more
accurate estimation of surface orientation, compared to using
only isotropic receptive fields that do not support affine covariance
(Lindeberg and G{\aa}rding \cite{LG96-IVC}). 

To implement affine covariance in a vision system corresponds to
using receptive field shapes subject to different affine spatial
transformations, and will thereby specifically imply that receptive
fields ought to be present for different degrees of elongation.%
\footnote{For a visual neuron with a spatial receptive field, one can
  conceive that the receptive field has different amounts of spatial extent in different
  spatial orientations in image space. One way to characterize the
  amount of elongation of a receptive field is by measuring the
  spatial extent over all possible orientations and then forming
  the ratio between the extreme values of the spatial extent over
  all the image orientations. For receptive field models formulated
  in terms of affine Gaussian derivatives, as used in this work,
  the receptive field can specifically be decomposed into a smoothing stage with an
  affine Gaussian kernel followed by spatial derivative
  computations. For that affine Gaussian derivative model, the degree of
  elongation $\kappa$ can be defined as the square root of the ratio of the
  eigenvalues of spatial covariance matrix $\Sigma$ in the affine Gaussian
  kernel according to Eq.~(\ref{eq-2D-aff-gauss}),
  as later formalized in Eq.~(\ref{eq-def-kappa}).}
In a companion work (Lindeberg \cite{Lin25-JCompNeurSci-spanelong}),
we have indeed investigated the consistency of that affine
covariant hypothesis with neurophysiological measurements of the resultant
of orientation selectivity curves obtained by
Goris {\em et al.\/}\ \cite{GorSimMov15-Neuron}.
There, we showed that predictions of orientation selectivity curves
and orientation selectivity histograms generated from idealized models
of simple cells in terms of Gaussian derivatives are for idealized
model of simple cells in reasonable agreement with biological
orientation selectivity histograms. Thereby, those results for simple
cells are consistent with an
expansion of the receptive field shapes over the degree of elongation
in the primary visual cortex of higher mammals.

The modelling of complex cells%
\footnote{According to the taxonomy of neurons in the primary visual
cortex by Hubel and Wiesel
\cite{HubWie59-Phys,HubWie62-Phys,HubWie68-JPhys,HubWie05-book},
a visual neuron is said to be simple if it (i) has distinct
excitatory and inhibitory subregions, (ii) obeys roughly linear
summation properties and (iii) the excitatory
and inhibitory regions balance each other in diffuse lighting.
A visual neuron that does not obey these properties is said to
be a complex cell. This characterization applies to any single individual visual
neuron. In our comparisons to biological data to be performed later
in this paper, we will then compare statistics over populations of
biological complex cells, for relating gross properties of our proposed
computational models to biological data.}
performed in
Lindeberg \cite{Lin25-JCompNeurSci-spanelong} was, however,
based on very much simplified models in terms of {\em pointwise\/} non-linear
combinations of responses of simple cells, and thus not involving any
spatial integration of the non-linear combinations over extended
regions in the image domain. Additionally, the modelling of complex
cells in Lindeberg \cite{Lin25-JCompNeurSci-spanelong} was
based on input from simple cells up to only order two, and not
involving simple cells up to order 4, which were shown to lead to better agreement
between the predicted orientation selectivity histograms
and the actual biological orientation selectivity histograms
accumulated for simple cells by Goris {\em et al.\/}\ \cite{GorSimMov15-Neuron}.
The subject of this paper is to investigate a set of extended models
of complex cells, and to demonstrate that these models offer a
potential to lead to better agreement with biological orientation selectivity
histograms compared to previous work.

In this way, we will thus specifically demonstrate that the resulting
modelling of complex cells is also consistent with an expansion of
the receptive field shapes over the degree of elongation, and in
this way consistent with the wider hypothesis about affine covariant
visual receptive fields, previously proposed in
Lindeberg
\cite{Lin21-Heliyon,Lin23-FrontCompNeuroSci,Lin25-JCompNeurSci-spanelong}.
Based on these results, to be presented below, we propose to include an explicit
expansion over the degree of elongation of the receptive fields as an
essential component when modelling the computational function of
complex cells.

While the specific models of complex cells to be considered in this
treatment will be highly idealized, in terms of generalized quadratic energy
models of the output of idealized simple cells, we propose that the
conceptual extension of models of complex cells with regard to an
expansion over the degree of elongation of the receptive fields ought
to generalize to also other types of functional models of complex cells.

More generally, we propose that the presented methodology
made use of in this paper, of subjecting computational models
of complex cells to similar probing tests as used for probing
the orientation selectivity properties of biological complex cells,
is important for understanding the theoretical properties of
the computational mechanisms that are used for modelling
biological complex cells. In this way, we will specifically
demonstrate that complementing previous energy models
of complex cells based on Gaussian derivatives with a
spatial integration stage, as well as making use of
spatial derivatives up to order 4 as opposed to previous
use of spatial derivatives only up to order 2, has the
potential of leading to better explanatory properties
of orientation selectivity histograms of biological neurons,
compared to the previous modelling approaches
in Lindeberg \cite{Lin25-JCompNeurSci-orisel,Lin25-JCompNeurSci-spanelong}.

Another important aspect of the presented work is that the proposed
models for complex cells are based on theoretically well-founded
models of simple cells, and specifically with very few free parameters
to determine.
By this {\em functional\/} modelling of the receptive fields at a coarse
{\em macroscopic\/} level, the need for determining hyperparameters
of the models is far lower compared what would be the case
if one would instead base the analysis on more fine-grained
models based on explicit neural models.

Thereby, the simulation work needed to reveal the qualitative
properties of the computational models also becomes far lower compared
to explicit simulations of networks of neural models, for which
the result may also depend on the settings of the hyperparameters,
and for which there may not be sufficient neurophysiological data
available to tune the hyperparameters in a well-founded manner.

A general biological motivation for this study is that biological experiments
often tend to reveal a variability of neurons in a variety of different respects.
Connectivity analysis of the anatomy also tend to reveal a convergence
along axonal projections.
In this treatment, for idealized models of complex cells,
we explain why there ought to be a variability in the degree of
elongation, because of
desirable affine covariance properties of an idealized vision system.
Specifically, we demonstrate that this hypothesis is consistent with experimental
results regarding orientation selectivity properties of biological neurons.

The theory of affine Gaussian derivative operators used as spatial
models of the receptive fields also gives a theoretical motivation
for the use of Gaussian derivative operators for different orders of
spatial differentiation, with different numbers of main lobes in
the receptive fields as function of the order of spatial differentiation.
In combination with orientation selectivity histograms over the
resultant of the orientation selectivity distributions for different
orders of spatial differentiation, we demonstrate that the formulation
of idealized models of complex cells based on a richer set of
spatial derivatives leads to more uniform orientation selectivity
histograms with closer similarity to orientation selectivity
histograms accumulated from biological neurons.

In these ways, we demonstrate that the proposed computational
mechanisms in terms of (i)~a variability over the degree of elongation
of the receptive fields, (ii)~the combination of receptive field
components corresponding to a richer set of orders of spatial
differentiation, and (iii)~the inclusion of explicit mechanisms for
spatial integration provide ways towards bridging the gap between
theoretical models of neural computation in relation to experimental
results obtained from neurophysiological recordings of biological
neurons.

Finally, we will use the results from the presented treatment 
(i)~for stating a
set of more general predictions in the section ``\nameref{sec-predictions}'',
to be used as guide for further modelling of
biological complex cells by mathematical models, and
(ii)~proposing a conceptual extension 
of the previous methodology for probing the orientation selectivity of
biological neurons.
The latter extension consists of instead of just recording the result for a single angular frequency
for each image orientation instead performing a two-parameter variation
over both the angular frequency and the image orientation, to be able
to better reflect a variability in the degree of elongation between
different individual biological neurons.

While the presented computational mechanisms are in the paper technically
developed for the proposed family of affine quasi quadrature models of
complex cells, we argue that corresponding additions of such
computational mechanisms could be considered also for other types of
theoretical and computational models of complex cells.

\section*{Methods}

\subsection*{Related work}

Orientation selectivity properties of biological neurons have been
studied by Watkins and Berkley \cite{WatBer73-ExpBrainRes},  
Rose and Blakemore \cite{RosBla74-ExpBrainRes},
Schiller {\em et al.\/} \cite{SchFinVol76-JNeuroPhys},
Albright \cite{Alb84-JNeuroPhys},
Ringach {\em et al.\/} \cite{RinShaHaw03-JNeurSci},
Nauhaus {\em et al.\/}\ \cite{NauBenCarRin09-Neuron},
Scholl {\em et al.\/} \cite{SchTanCorPri13-JNeurSci},
Sadeh and Rotter \cite{SadRot14-BICY},
Goris {\em et al.\/}\ \cite{GorSimMov15-Neuron}
and
Sasaki {\em et al.\/}
\cite{SakKimNimTabTanFukAsaAraInaNakBabDaiNisSanTanImaTanOhz15-SciRep}.
Biological mechanisms for achieving orientation selectivity have
also been investigated by
Somers {\em et al.\/}\ \cite{SomNelSur95-JNeuroSci},
Sompolinsky and Shapley \cite{SomSha97-CurrOpNeuroBio},
Carandini and Ringach \cite{CarRin97-VisRes},
Lampl {\em et al.\/}\ \cite{LamAndGilFer01-Neuron},
Ferster and Miller \cite{FerMil00-AnnRevNeuroSci},
Shapley {\em et al.\/}\ \cite{ShaHawRin03-Neuron},
Seri{\`e}s {\em et al.\/}\ \cite{SerLatPou04-NatNeuroSci},
Hansel and van~Vreeswijk \cite{HanVre12-JNeuroSci},
Moldakarimov {\em et al.\/}\ \cite{MolBazSej14-PLOSCompBiol},
Gonzalo Cogno and Mato \cite{GonMat15-FrontNeurCirc},
Priebe \cite{Pri16-AnnRevVisSci},
Pattadkal {\em et al.\/} \cite{PatMatVrePriHan18-CellRep},
Nguyen and Freeman \cite{NguFre19-PLOSCompBiol},
Merkt {\em et al.\/} \cite{MerSchRot19-PLOSCompBiol},
Wei {\em et al.\/} \cite{WeiMerRot22-bioRxiv}
and
Wang {\em et al.\/} \cite{WanDeyLagBehCalSta24-CellRep}.
In this paper, our focus is, however, not on
neural mechanisms, but on {\em functional properties\/}
at a macroscopic level.

Receptive field models of simple in terms of Gaussian derivatives
have been formulated by
Koenderink and van Doorn \cite{Koe84,KoeDoo87-BC,KoeDoo92-PAMI},
Young and his co-workers
\cite{You87-SV,YouLesMey01-SV,YouLes01-SV}
and Lindeberg \cite{Lin13-BICY,Lin21-Heliyon},
and in terms of Gabor functions by
Marcelja \cite{Mar80-JOSA},
Jones and Palmer \cite{JonPal87a,JonPal87b} and
Porat and Zeevi \cite{PorZee88-PAMI}.
More extensive theoretical models based on Gaussian derivatives
have also been expressed by
Lowe \cite{Low00-BIO},
May and Georgeson \cite{MayGeo05-VisRes},
Hesse and Georgeson \cite{HesGeo05-VisRes},
Georgeson  {\em et al.\/}\ \cite{GeoMayFreHes07-JVis},
Hansen and Neumann \cite{HanNeu09-JVis},
Wallis and Georgeson \cite{WalGeo09-VisRes},  
Wang and Spratling \cite{WanSpra16-CognComp},
Pei {\em et al.\/}\ \cite{PeiGaoHaoQiaAi16-NeurRegen},
Ghodrati {\em et al.\/}\ \cite{GhoKhaLeh17-ProNeurobiol},
Kristensen and Sandberg \cite{KriSan21-SciRep},
Abballe and Asari \cite{AbbAsa22-PONE},
Ruslim {\em et al.\/}\ \cite{RusBurLia23-bioRxiv} and
Wendt and Faul \cite{WenFay24-JVis}.

The taxonomy into simple and complex cells in the primary visual
cortex was proposed in the pioneering work by
Hubel and Wiesel
\cite{HubWie59-Phys,HubWie62-Phys,HubWie68-JPhys,HubWie05-book}.
More extensive analysis of properties of simple cells have
then been presented by
DeAngelis {\em et al.\/}\
\cite{DeAngOhzFre95-TINS,deAngAnz04-VisNeuroSci},
Ringach \cite{Rin01-JNeuroPhys,Rin04-JPhys},
Conway and Livingstone \cite{ConLiv06-JNeurSci},
Johnson {\em et al.\/}\ \cite{JohHawSha08-JNeuroSci},
Walker {\em et al.\/}
\cite{WalSinCobMuhFroFahEckReiPitTol19-NatNeurSci}
and
De and Horwitz \cite{DeHor21-JNPhys},
and regarding complex cells by
Movshon {\em et al.\/}\ \cite{MovThoTol78-JPhys}, 
Emerson {\em et al.\/}\ \cite{EmeCitVauKle87-JNeuroPhys},
Martinez and Alonso \cite{MarAlo01-Neuron},
Touryan {\em et al.\/}\ \cite{TouLauDan02-JNeuroSci,TouFelDan05-Neuron},
Rust {\em et al.\/}\ \cite{RusSchMovSim05-Neuron},
van~Kleef {\em et al.\/}\ \cite{KleCloIbb10-JPhys},  
Goris {\em et al.\/}\ \cite{GorSimMov15-Neuron},
Li {\em et al.\/}\ \cite{LiLiuChoZhaTao15-JNeuroSci} and
Almasi {\em et al.\/}\ \cite{AlmMefCloWonYunIbb20-CerCort},
as well as modelled computationally by
Adelson and Bergen \cite{AdeBer85-JOSA},
Heeger \cite{Hee92-VisNeuroSci},
Serre and Riesenhuber \cite{SerRie04-AIMemo},
Einh{\"a}user {\em et  al.\/} \cite{EinKayKoeKoe02-EurJNeurSci},
Kording {\em et al.\/}\ \cite{KorKayWinKon04-JNeuroPhys},
Merolla and Boahen \cite{MerBoa04-NIPS},
Berkes and Wiscott \cite{BerWis05-JVis},
Carandini \cite{Car06-JPhys},
Hansard and Horaud \cite{HanHor11-NeurComp},
Franciosini {\em et al.\/}\ \cite{FraBouPer19-AnnCompNeurSciMeet},
Lindeberg \cite{Lin20-JMIV},
Lian {\em et  al.\/} \cite{LiaAlmGraKamBurMef21-PLOSCompBiol},
Oleskiw {\em et al.\/}\ \cite{OleLieSimMov23-bioRxiv},
Yedjour and Yedjour \cite{YedYed24-CognNeurDyn},
Nguyen {\em et al.\/}\ \cite{NguSooHuaBak24-PLOSCompBio} and
Almasi {\em et al.\/}\ \cite{AlmSunJunIbbMef25-JNeurEng}.

Notably, in relation to the generalized quadratic models of complex
cells in V1 to be considered in this paper,
Rowekamp and Sharpee \cite{RowSha25-PLOSCompBiol}
have found that quadratic computations strongly increase both the predictive power
of their models of visual neurons in V1, V2 and V4 as well
as their neural selectivity to natural stimuli.

There have been some neurophysiological studies reported that mention
receptive fields with different aspect ratios
(Tinsley {\em et al.\/}\ \cite{TinWebBarVinParDer03-JNeuroPhys},
Xu  {\em et al.\/}\ \cite{XuLiCheWanYan16-CognNeurDyn}).
According to our knowledge, there have, however, not been any
previously developed models of complex cells, that involve
explicit expansions of the receptive field shapes
over the degree of elongation of the receptive fields,
or or more specifically that are able to match the neurophysiological
orientation selectivity histograms accumulated by
Goris  {\em et al.\/}\ \cite{GorSimMov15-Neuron}.

\subsection*{Background theory}

In this section, we will describe basic properties of the idealized
models for idealized models of receptive fields and their orientation
selectivity properties, which we will then build upon and
extend in the section '\nameref{sec-results}''.

\subsubsection*{Idealized models for spatial receptive fields}

For modelling simple and complex cells in the primary visual cortex,
we will build upon the generalized Gaussian derivative
model for visual receptive fields proposed in
Lindeberg \cite{Lin13-BICY,Lin21-Heliyon}
and further developed in
Lindeberg
\cite{Lin23-FrontCompNeuroSci,Lin25-JCompNeurSci-orisel,Lin25-JCompNeurSci-spanelong,Lin25-JMIV}.

\paragraph{Models for simple cells.}

According to this theory, linear models of purely spatial receptive fields
corresponding to simple cells are formulated in terms of affine Gaussian
derivatives of the form
\begin{multline}
  \label{eq-spat-RF-model}
  T_{\simple}(x_1, x_2;\; \sigma_{\varphi}, \varphi, \Sigma_{\varphi}, m) = \\
  = T_{\varphi^m,\norm}(x_1, x_2;\; \sigma_{\varphi}, \Sigma_{\varphi})
  = \sigma_{\varphi}^{m} \, \partial_{\varphi}^{m} \left( g(x_1, x_2;\; \Sigma_{\varphi}) \right),
\end{multline}
where
\begin{itemize}
\item
   $\varphi \in [-\pi, \pi]$ is the preferred orientation of the receptive
   field,
\item
  $\sigma_{\varphi} \in \bbbr_+$ is the amount of spatial smoothing,
\item
  $\partial_{\varphi}^m =
  (\cos \varphi \, \partial_{x_1} + \sin  \varphi \, \partial_{x_2})^m$
  is an $m$:th-order directional derivative operator,
   in the direction $\varphi$,
 \item
   $\Sigma_{\varphi}$ is a $2 \times 2$ symmetric positive definite covariance matrix, with
   one of its eigenvectors in the direction of $\varphi$, 
 \item
   $g(x;\; \Sigma_{\varphi})$ is a 2-D affine Gaussian kernel with its shape
   determined by the spatial covariance matrix $\Sigma_{\varphi}$
   \begin{equation}
     \label{eq-2D-aff-gauss}
     g(x;\; \Sigma_{\varphi})
     = \frac{1}{2 \pi \sqrt{\det \Sigma_{\varphi}}}
         e^{-x^T \Sigma_{\varphi}^{-1} x/2}
    \end{equation}
    for $x = (x_1, x_2)^T \in \bbbr^2$.
\end{itemize}
In Lindeberg \cite{Lin21-Heliyon}, it was demonstrated
that idealized receptive field models of this type do
rather well model the qualitative shape of biological
simple cells as obtained by neurophysiological measurements
by DeAngelis {\em et al.\/}\
\cite{DeAngOhzFre95-TINS,deAngAnz04-VisNeuroSci},
Conway and Livingstone \cite{ConLiv06-JNeurSci} and
Johnson {\em et al.\/}\ \cite{JohHawSha08-JNeuroSci}.

\begin{figure}[hbtp]
  \begin{center}
     \begin{tabular}{ccc}
     \includegraphics[width=0.30\textwidth]{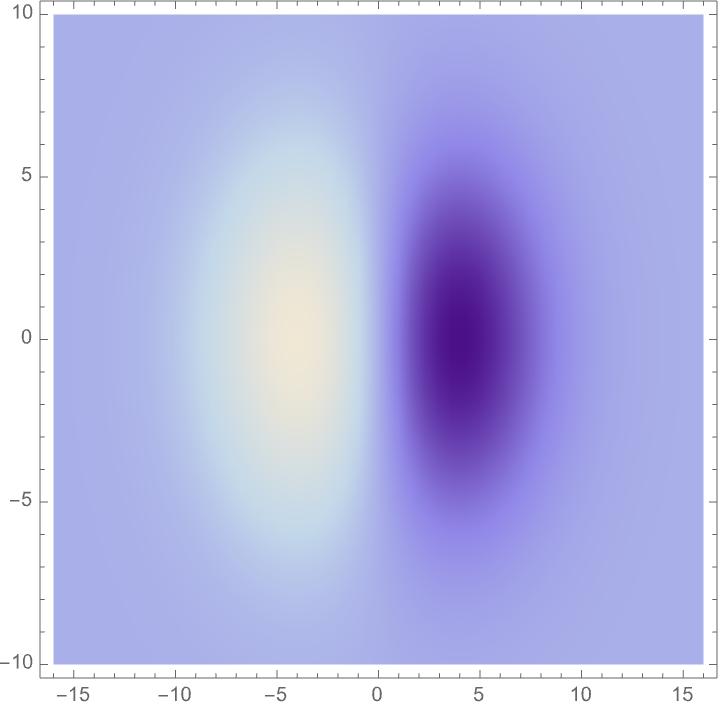}
       & \includegraphics[width=0.30\textwidth]{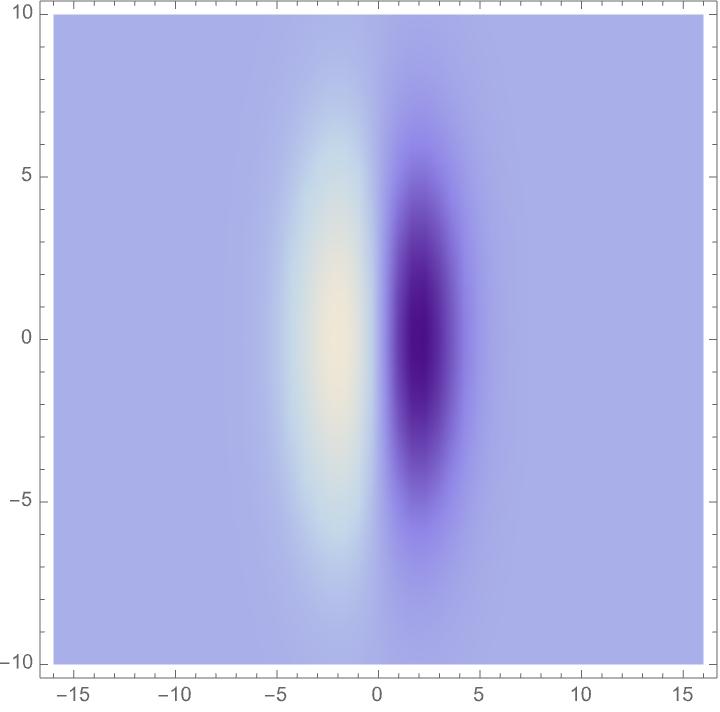}
       & \includegraphics[width=0.30\textwidth]{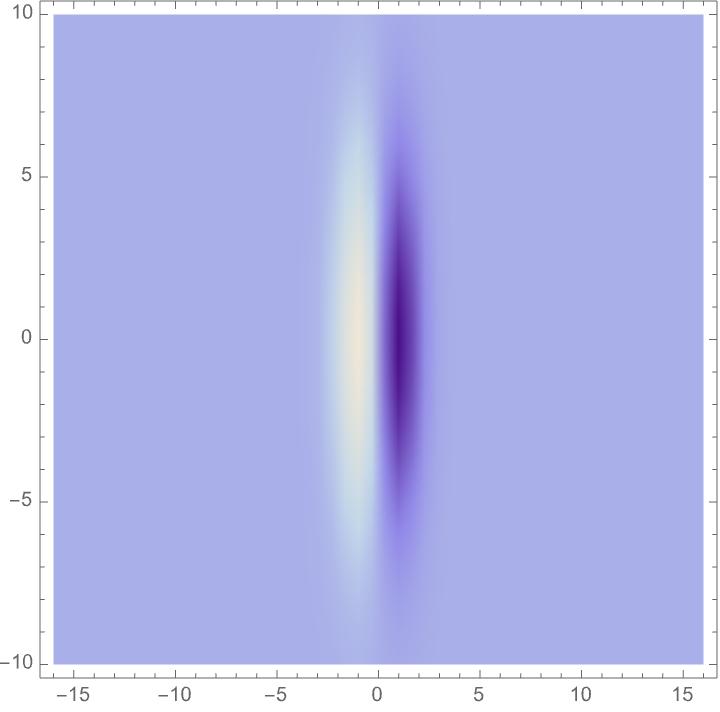} \\
     \includegraphics[width=0.30\textwidth]{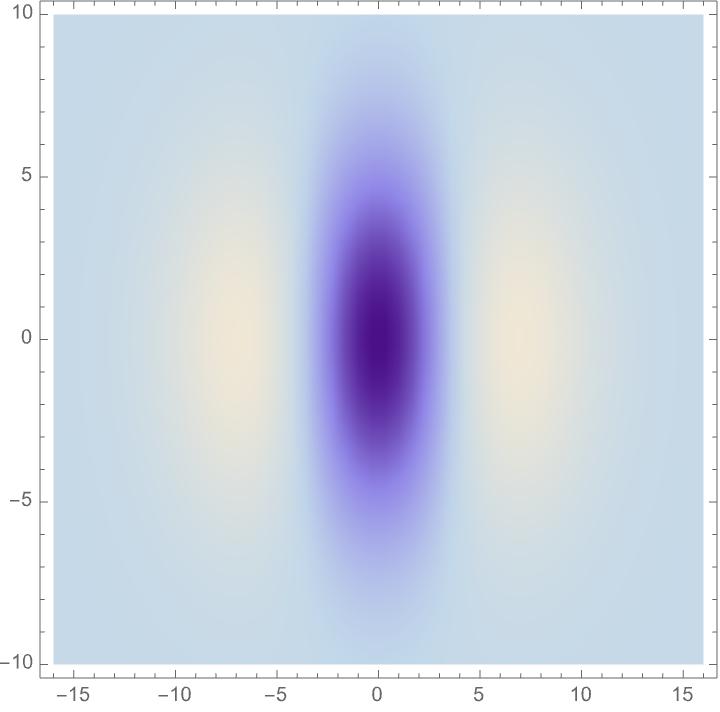}
       & \includegraphics[width=0.30\textwidth]{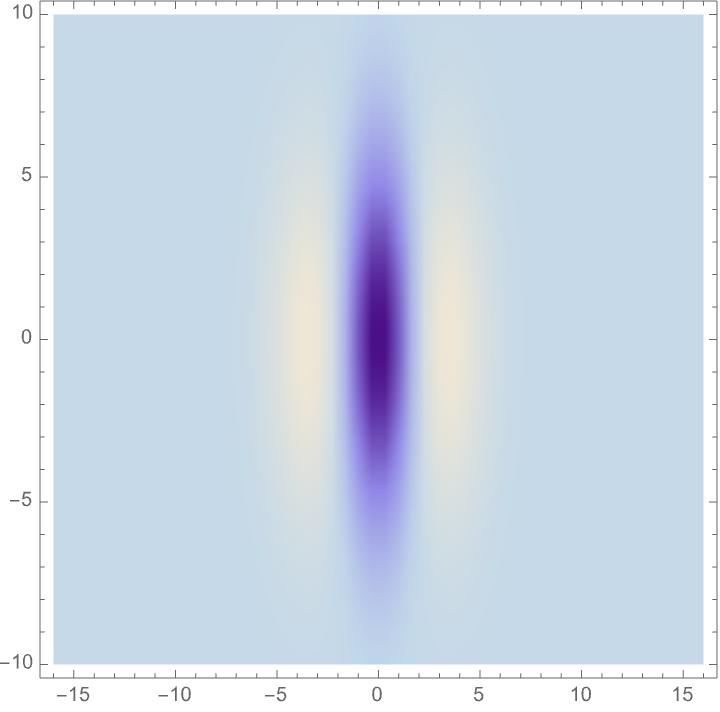}
       & \includegraphics[width=0.30\textwidth]{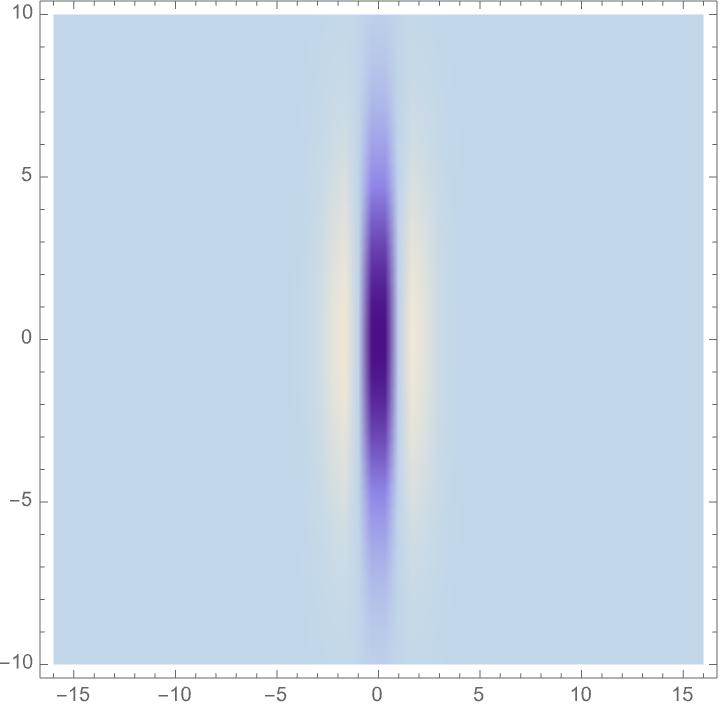} \\
     \includegraphics[width=0.30\textwidth]{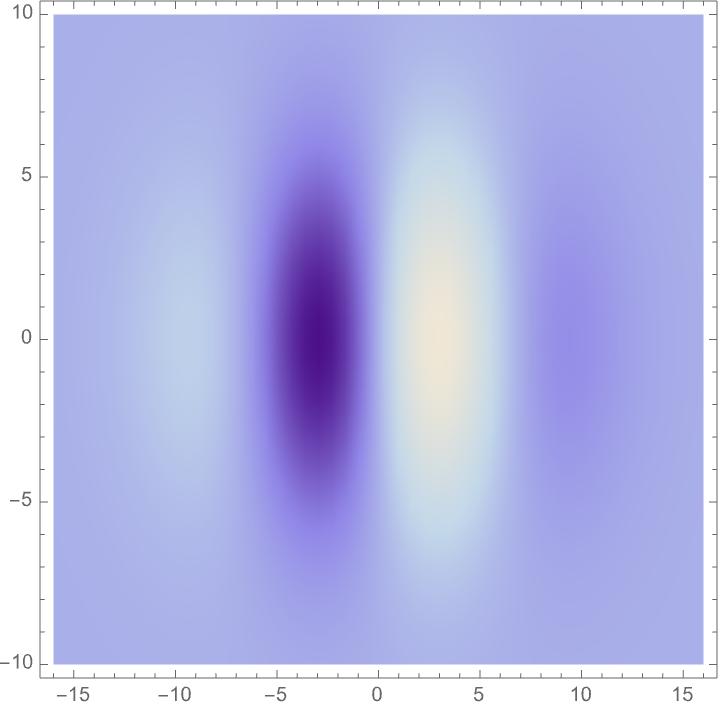}
       & \includegraphics[width=0.30\textwidth]{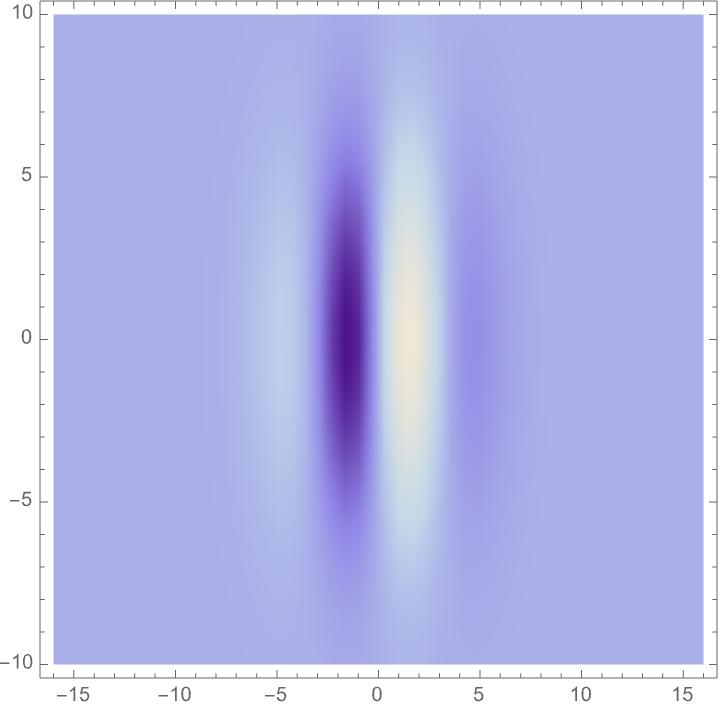}
       & \includegraphics[width=0.30\textwidth]{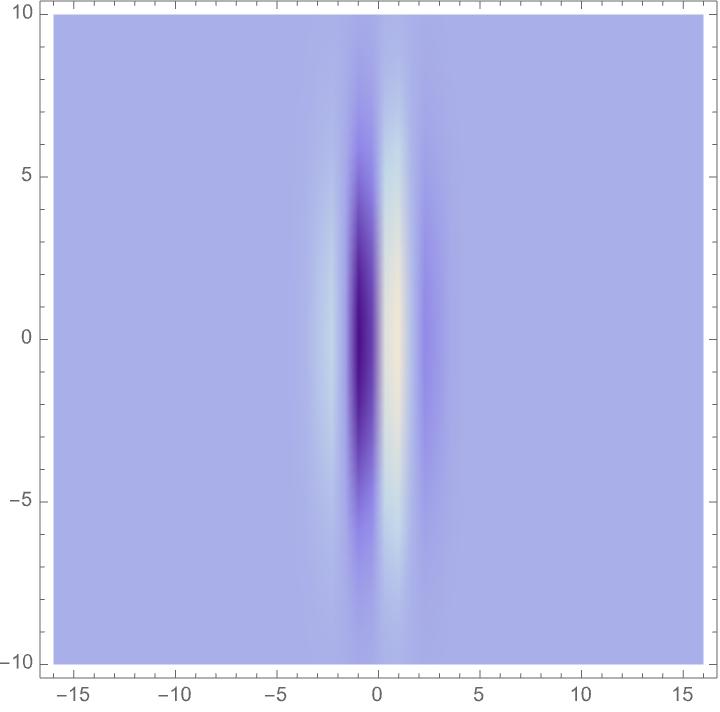} \\
     \includegraphics[width=0.30\textwidth]{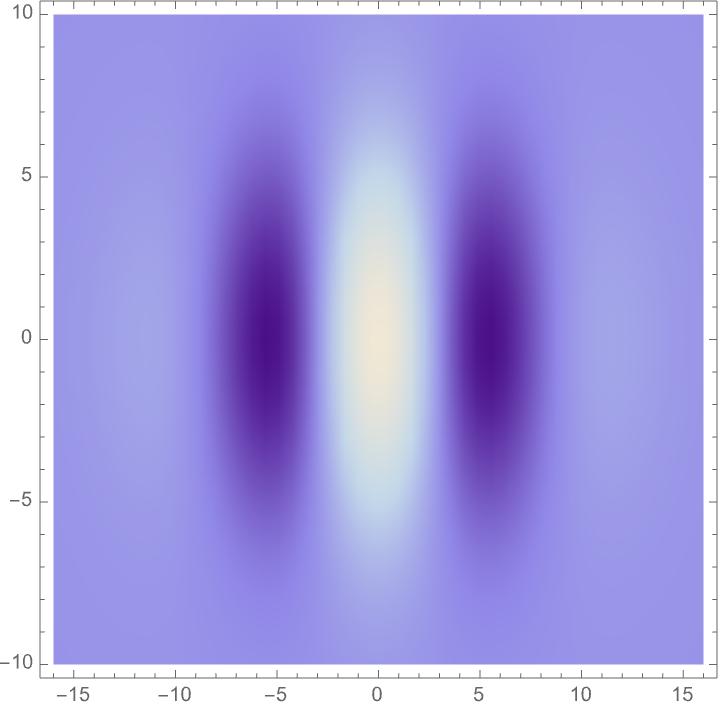}
       & \includegraphics[width=0.30\textwidth]{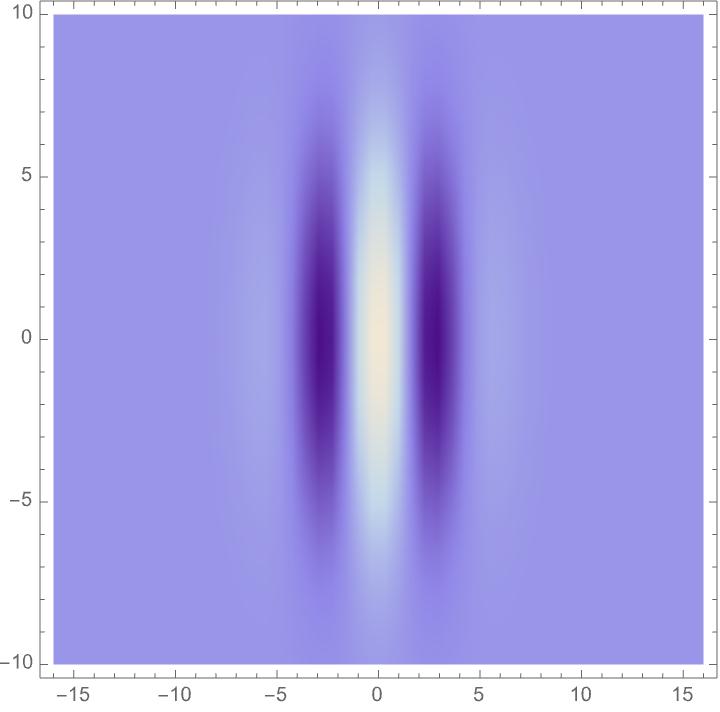}
       & \includegraphics[width=0.30\textwidth]{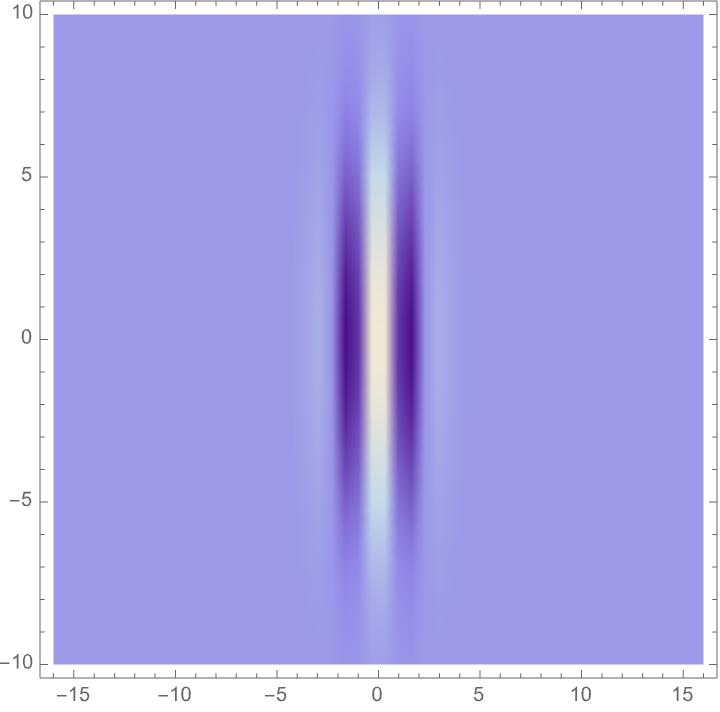} \\
     \end{tabular}
  \end{center}
  \caption{{\bf Distribution of affine Gaussian derivative
    receptive fields} (for the preferred image orientation $\varphi = 0$)
    over the scale parameter ratio 
    $\kappa = \sigma_2/\sigma_1$ between $1$ to $4$,
    from left to right.
    Here, the vertical scale parameter kept is constant $\sigma_2 = 4$,
    while the horizontal scale parameter is smaller
    $\sigma_1 \leq \sigma_2$.
    (first row) First-order directional derivatives according
    to (\ref{eq-spat-RF-model}) for $m = 1$.
    (second row) Second-order directional derivatives
    according to (\ref{eq-spat-RF-model}) for $m = 2$.
    (third row) Third-order directional derivatives
    according to (\ref{eq-spat-RF-model}) for $m = 3$.
    (fourth row) Fourth-order directional derivatives
    according to (\ref{eq-spat-RF-model}) for $m = 4$.
    (Horizontal axes: image coordinate $x_1 \in [-16, 16]$.
     Vertical axes: image coordinate $x_2 \in [-16, 16]$.)}
  \label{fig-ecc-variability}
\end{figure}

Fig.~\ref{fig-ecc-variability} shows examples of such
receptive fields for different orders of spatial differentiation
$m \in \{ 1, 2, 3, 4 \}$ and different values of the scale
parameter ratio $\kappa = \sigma_2/\sigma_1 \in \{ 1, 2, 4 \}$
between the here vertical and the horizontal scale parameters
$\sigma_2$ and $\sigma_1$, respectively, for the here preferred
orientation $\varphi = 0$ for the receptive fields. In addition to
this illustrated variability over the degree of elongation, an
idealized vision system should additionally comprise a variability
over the preferred orientation $\varphi$ of the receptive fields,
and possibly also over the overall size of the receptive fields,
as further developed in Lindeberg \cite{Lin25-JCompNeurSci-spanelong}.

\paragraph{Models for complex cells.}

As a simplest possible extension to non-linear complex cells,
an affine quasi-quadrature measure of the form (Lindeberg
\cite{Lin20-JMIV} Eq.~(39))
\begin{equation}
  \label{eq-quasi-quad-dir}
  {\cal Q}_{\varphi,12,\text{pt}} L
  = \sqrt{L_{\varphi,\norm}^2+ C_{\varphi} \, L_{\varphi\varphi,\norm}^2},
\end{equation}
was studied in Lindeberg \cite{Lin25-JCompNeurSci-spanelong},
where
\begin{itemize}
\item
  $L_{\varphi,\norm}$ and $L_{\varphi\varphi,\norm}$ denote
  directional derivatives in the direction $\varphi$
  of orders 1 and~2 of convolutions of the input image $f(x_1, x_2)$
  with affine Gaussian derivative kernels of the form (\ref{eq-spat-RF-model}):
  \begin{equation}
    \label{eq-def-Lphi}
       L_{\varphi,\norm}(x_1, x_2;\;  \sigma_{\varphi}, \Sigma_{\varphi}) 
       = T_{\varphi,\norm}(x_1, x_2;\; \sigma_{\varphi}, \Sigma_{\varphi}) *
       f(x_1, x_2),
    \end{equation}
    \begin{equation}
      \label{eq-def-Lphiphi}
       L_{\varphi\varphi,\norm}(x_1, x_2;\;  \sigma_{\varphi}, \Sigma_{\varphi}) 
        = T_{\varphi\varphi,\norm}(x_1, x_2;\; \sigma_{\varphi}, \Sigma_{\varphi}) *
       f(x_1, x_2),
    \end{equation}
\item
  $C_{\varphi} > 0$ is a weighting factor between first and second-order
  information, which based on a theoretical analysis in
  Lindeberg \cite{Lin18-SIIMS} is often set to $C = 1/\sqrt{2}$.
\end{itemize}
This model is closely related to the energy model of complex cells
proposed by Adelson and Bergen \cite{AdeBer85-JOSA} and
Heeger \cite{Hee92-VisNeuroSci}, as well as inspired by
the fact that odd- and even-shaped receptive fields have
been reported to occur in pairs
 (De~Valois {\em et al.\/}\ \cite{ValCotMahElfWil00-VR}).
The quasi quadrature serves as an approximation of a quadrature
pair, as formulated based on a Hilbert transform
(Bracewell \cite{Bra99}, pp.~267--272), although
instead formulated in terms of affine Gaussian derivatives,
which are then summed up in squares in to reduce the
phase dependency; see Lindeberg \cite{Lin20-JMIV}
for further details.

Concerning the validity of such an energy-based model of simple cells
to model the computational function of complex cells, it is
interesting to note that when
Touryan {\em et al.\/} \cite{TouFelDan05-Neuron}
extracted the eigenvectors of second-order Wiener kernels
to model the computational function of complex cells,
the first two eigenvectors turned out the have spatial shapes that
very well agree with the shapes of first- and second-order Gaussian
derivatives;
compare with Fig.~5B in Touryan {\em et al.\/} \cite{TouFelDan05-Neuron}.
Hence, even though an energy model of complex cells in terms of the
output from a set of simple cells may not be able to span the full flexibility in terms of
the possible computational functions of biological complex cells, an
approximation of the computational function of a complex cell in terms
of an energy model in such a way ought to be able to reveal essential properties of
the computational functionalities of complex cells.

\subsubsection*{Orientation selectivity properties}

\begin{figure}[hbtp]
  \begin{center}
    \includegraphics[width=0.60\textwidth]{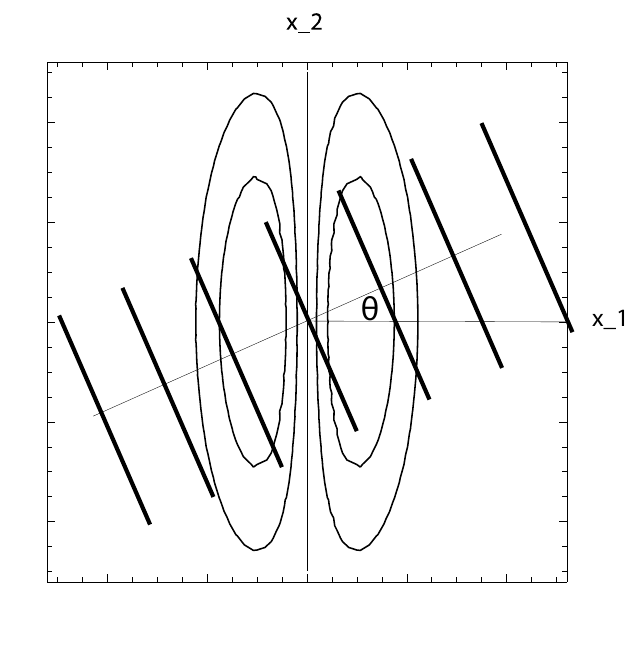}
  \end{center}
  \caption{{\bf Configuration for probing the orientation selectivity
      of an idealized receptive field model.} To measure the orientation selectivity properties of
    an idealized receptive field (here illustrated by the level curves
    of a first-order derivative of an affine Gaussian kernel in
    the horizontal direction with preferred orientation $\varphi = 0$),
    we compute the response to a sine wave (here illustrated by a
    set of darker level lines corresponding to the spatial maxima
    and minima of the sine wave) with inclination angle $\theta$.
    (Horizontal axis: spatial coordinate $x_1$.
    Vertical axis: spatial coordinate $x_2$.)
  (Adapted from Lindeberg
  \cite{Lin25-JCompNeurSci-orisel}
  OpenAccess.)}
  \label{fig-schem-ill-model-rf-sine}
\end{figure}

In Lindeberg
\cite{Lin25-JCompNeurSci-orisel,Lin25-JCompNeurSci-spanelong},
the orientation
selectivity properties of these idealized models of simple and complex
cells were investigated in detail, by computing the responses to
sine wave functions of the form
\begin{equation}
  \label{eq-sine-wave-model-spat-anal}
  f(x_1, x_2) =
  \sin
  \left(
    \omega \cos (\theta) \, x_1 + \omega \sin (\theta) \, x_2+ \beta
  \right)
\end{equation}
where $\theta \in [-\pi/2, \pi/2]$ denotes the inclination angle of the sine wave in
relation to the preferred orientation $\varphi = 0$ of the receptive field and
$\beta$ denotes the phase;
see Fig.~\ref{fig-schem-ill-model-rf-sine}
for an illustration.

Specifically, in Lindeberg
\cite{Lin25-JCompNeurSci-orisel,Lin25-JCompNeurSci-spanelong},
it was shown that with 
\begin{equation}
  \label{eq-def-kappa}
  \kappa = \frac{\sigma_2}{\sigma_1}
\end{equation}
denoting the ratio between the scale parameters $\sigma_1 \in \bbbr_+$ and
$\sigma_2 \in \bbbr_+$ of the affine Gaussian kernel in the preferred
direction {\em vs.\/}\ the orthogonal direction of the receptive
field, the orientation selectivity curves are of the form
%
\begin{equation}
  \label{eq-ori-sel-curves-gen-gauss-der-model-general}
  r_{\lambda}(\theta)
  = \left(
         \frac{\left| \cos \theta \right|}
         {\sqrt{\cos ^2 \theta + \kappa ^2 \sin ^2\theta}}
         \right)^{\lambda}
\end{equation}
with
\begin{itemize}
\item
  $\lambda = m$ for an $m$:th-order model of a simple
cell of the form (\ref{eq-spat-RF-model}) for
$m \in \{ 1, 2, 3, 4 \}$ and
\item
  $\lambda = 3/2$
for an idealized model of a complex cell of the form
(\ref{eq-quasi-quad-dir}).
\end{itemize}
Notably, for all these idealized models of the receptive fields,
a smaller value of the scale parameter ratio $\kappa$
leads to wider orientation selectivity curves, whereas larger
values of $\kappa$ lead to more narrow orientation selectivity
properties. In this respect, given the assumption that the
affine Gaussian derivative model should constitute an appropriate
model for the spatial component of simple cells,
a variability in the degree of elongation of the receptive fields
will correspond to a variability in the orientation selectivity
and {\em vice versa\/}.

\section*{Results}
\label{sec-results}

\subsection*{Generalized idealized models of complex cells}

In this work, we will extend the idealized model
(\ref{eq-quasi-quad-dir}) for complex cells in two major ways:
\begin{itemize}
\item
  by adding a spatial integration stage, and
\item
  considering derivatives up to order 4 in addition to derivatives
  up to order 2.
\end{itemize}
The motivation for adding a spatial integration stage
is that the model for complex cell should then make use of input
from more than two simple cells, specifically by accumulating
input over multiple positions in the visual field.

The motivation for adding derivatives
of higher order than 2 are: (i)~to enable more narrow orientation
selectivity properties and (ii)~in Lindeberg
\cite{Lin25-JCompNeurSci-spanelong} it was shown that models of
simple cells up to order 4 lead to better agreement with the
orientation selectivity histogram of simple cells recorded by
Goris {\em et al.\/}\ \cite{GorSimMov15-Neuron} than
simple cells up to order 2.

Thus, if we let $L_{\varphi\varphi\varphi,\norm}$ and
$L_{\varphi\varphi\varphi\varphi,\norm}$ denote
the directional derivatives in the direction $\varphi$
of orders 3 and~4 of convolutions of the input image $f(x_1, x_2)$
with affine Gaussian derivative kernels of the form (\ref{eq-spat-RF-model}):
\begin{equation}
  \label{eq-def-Lphiphiphi}
  L_{\varphi\varphi\varphi,\norm}(x_1, x_2;\;  \sigma_{\varphi}, \Sigma_{\varphi}) 
  = T_{\varphi\varphi\varphi,\norm}(x_1, x_2;\; \sigma_{\varphi}, \Sigma_{\varphi}) *
  f(x_1, x_2),
\end{equation}
\begin{equation}
  \label{eq-def-Lphiphiphiphi}
  L_{\varphi\varphi\varphi\varphi,\norm}(x_1, x_2;\;  \sigma_{\varphi}, \Sigma_{\varphi}) 
  = T_{\varphi\varphi\varphi\varphi,\norm}(x_1, x_2;\; \sigma_{\varphi}, \Sigma_{\varphi}) *
  f(x_1, x_2),
\end{equation}
as well as consider the previous definitions of
$L_{\varphi,\norm}$ and
$L_{\varphi\varphi,\norm}$ according to
(\ref{eq-def-Lphi}) and (\ref{eq-def-Lphiphi}),
we will consider the following new generalized and integrated affine quasi
quadrature measures for modelling complex cells:
\begin{equation}
  \label{eq-int-quasi-quad12-dir}
    {\cal Q}_{\varphi,12,\text{int}} L 
    = \sqrt{\sum_{m \in \{ 1, 2 \}} g(\cdot, \cdot;\; \gamma^2 \, \Sigma_{\varphi})
      * C_{\varphi}^{m-1} \,  L_{\varphi^m,\norm}^2(\cdot, \cdot;\; \sigma_{\varphi}, \Sigma_{\varphi})}
\end{equation}
\begin{equation}
  \label{eq-int-quasi-quad1234-dir}
    {\cal Q}_{\varphi,1234,\text{int}} L 
    = \sqrt{\sum_{m \in \{ 1, 2, 3, 4 \}} g(\cdot, \cdot;\; \gamma^2 \, \Sigma_{\varphi})
      * C_{\varphi}^{m-1} \,  L_{\varphi^m,\norm}^2(\cdot, \cdot;\; \sigma_{\varphi}, \Sigma_{\varphi})}
\end{equation}
\begin{equation}
  \label{eq-int-quasi-quad34-dir}
    {\cal Q}_{\varphi,34,\text{int}} L 
    = \sqrt{\sum_{m \in \{ 3, 4 \}} g(\cdot, \cdot;\; \gamma^2 \, \Sigma_{\varphi})
      * C_{\varphi}^{m-3} \,  L_{\varphi^m,\norm}^2(\cdot, \cdot;\; \sigma_{\varphi}, \Sigma_{\varphi})},
\end{equation}
where $g(\cdot, \cdot;\; \gamma^2 \, \Sigma_{\varphi})$ for the relative
integration scale $\gamma > 1 $ denotes a spatially larger affine
Gaussian kernel than the affine Gaussian kernel
$g(\cdot, \cdot;\; \Sigma_{\varphi})$ used for computing
the receptive field responses  $L_{\varphi^m,\norm}$
for the idealized models of simple cells.
In the following experiments to be reported, we will throughout
use the parameter setting $\gamma = 1/\sqrt{2}$.

Structurally, these expressions are similar in the sense that the
squares of the directional derivative responses $L_{\varphi^m,\norm}$
are first integrated for different orders $m$ of spatial
differentiation, and then summed of for different subsets
$m \in \{ 1, 2 \}$, $m \in \{ 1, 2, 3, 4 \}$ and $m \in \{ 3, 4 \}$,
respectively.
Specifically, the first of these integrated affine quasi quadrature measures
${\cal Q}_{\varphi,12,\text{int}} L$ can basically be seen as a
spatially integrated extension of the pointwise affine quasi quadrature
measure ${\cal Q}_{\varphi,12,\text{pt}} L$ in (\ref{eq-quasi-quad-dir}).

\subsection*{Orientation selectivity curves for the generalized
  idealized models of complex cells}

To characterize the orientation selectivity properties for the
generalized integrated affine quasi quadrature measures
(\ref{eq-int-quasi-quad12-dir}), (\ref{eq-int-quasi-quad1234-dir})
and (\ref{eq-int-quasi-quad34-dir}), let us first compute the
responses $L_{\varphi^m,\norm}$ for the underlying models of simple
cells $T_{\varphi^m,\norm}$ to a sine wave
(\ref{eq-sine-wave-model-spat-anal}) 
according to Eqs.~(29) and~(35)
in Lindeberg \cite{Lin25-JCompNeurSci-orisel}
\begin{align}
   \begin{split}
     L_{0,\norm}(x_1, x_2;\; \sigma_1, \sigma_2) =
  \end{split}\nonumber\\
  \begin{split}
    & = \int_{\xi_1 = -\infty}^{\infty}  \int_{\xi_2 = -\infty}^{\infty}
             T_{0,\norm}(\xi_1, \xi_2;\; \sigma_1, \sigma_2) \,
              f(x_1 - \xi_1, x_2 - \xi_2) \, d \xi_1 \, d\xi_2
  \end{split}\nonumber\\
  \begin{split}
    & = \omega \, \sigma_1 \cos (\theta) \,
           e^{-\frac{1}{2} \omega^2 (\sigma_1^2 \cos^2 \theta + \sigma_2^2 \sin^2 \theta)}
    \label{eq-L0-pure-spat-anal}
           \, \cos
             (
                \omega \cos (\theta) \, x_1 + \omega \sin (\theta) \, x_2+ \beta
             ),
   \end{split}         
\end{align}
\begin{align}
   \begin{split}
     L_{00,\norm}(x_1, x_2;\; \sigma_1, \sigma_2) =
  \end{split}\nonumber\\
  \begin{split}
    & = \int_{\xi_1 = -\infty}^{\infty}  \int_{\xi_2 = -\infty}^{\infty}
             T_{00,\norm}(\xi_1, \xi_2;\; \sigma_1, \sigma_2) \,
             f(x_1 - \xi_1, x_2 - \xi_2) \, d \xi_1 \, d\xi_2
  \end{split}\nonumber\\
  \begin{split}
    & = - \omega^2 \, \sigma_1^2 \cos^2 (\theta) \,
           e^{-\frac{1}{2} \omega^2 (\sigma_1^2 \cos^2 \theta + \sigma_2^2 \sin^2 \theta)}
    \label{eq-L00-pure-spat-anal}
           \, \sin
             (
                \omega \cos (\theta) \, x_1 + \omega \sin (\theta) \, x_2 + \beta
             ),
   \end{split}         
\end{align}
and according to Eqs.~(36) and (42) in the supplementary material of
Lindeberg \cite{Lin25-JCompNeurSci-spanelong}
\begin{align}
   \begin{split}
     L_{000,\norm}(x_1, x_2;\; \sigma_1, \sigma_2) =
  \end{split}\nonumber\\
  \begin{split}
    & = \int_{\xi_1 = -\infty}^{\infty}  \int_{\xi_2 = -\infty}^{\infty}
             T_{000,\norm}(\xi_1, \xi_2;\; \sigma_1, \sigma_2) \,
              f(x_1 - \xi_1, x_2 - \xi_2) \, d \xi_1 \, d\xi_2
  \end{split}\nonumber\\
  \begin{split}
    & = - \omega^3 \, \sigma_1^3 \cos^3 (\theta) \,
           e^{-\frac{1}{2} \omega^2 (\sigma_1^2 \cos^2 \theta + \sigma_2^2 \sin^2 \theta)}
    \label{eq-L000-pure-spat-anal}
           \, \cos
             (
                \omega \cos (\theta) \, x_1 + \omega \sin (\theta) \, x_2 + \beta
             ),
   \end{split}         
\end{align}
\begin{align}
   \begin{split}
     L_{0000,\norm}(x_1, x_2;\; \sigma_1, \sigma_2) =
  \end{split}\nonumber\\
  \begin{split}
    & = \int_{\xi_1 = -\infty}^{\infty}  \int_{\xi_2 = -\infty}^{\infty}
             T_{0000,\norm}(\xi_1, \xi_2;\; \sigma_1, \sigma_2) \,
              f(x_1 - \xi_1, x_2 - \xi_2) \, d \xi_1 \, d\xi_2
  \end{split}\nonumber\\
  \begin{split}
    & = \omega^4 \, \sigma_1^4 \cos^4 (\theta) \,
           e^{-\frac{1}{2} \omega^2 (\sigma_1^2 \cos^2 \theta + \sigma_2^2 \sin^2 \theta)}
    \label{eq-L0000-pure-spat-anal}
           \, \sin
             (
                \omega \cos (\theta) \, x_1 + \omega \sin (\theta) \, x_2 + \beta
             ),
   \end{split}         
\end{align}
where $\sigma_1 \in \bbbr_+$ and $\sigma_2 \in \bbbr_+$
denote the scale parameters of the
affine Gaussian kernel in the horizontal and the vertical directions,
respectively.

Let us next integrate the squares of these expressions spatially using a Gaussian
window function with relative integration scale $\gamma > 1$, with the actual
calculations performed in Wolfram Mathematica
and leading to results that
are unfortunately too complex to be reproduced here.

Let us furthermore for the parameterization of the vertical scale
parameter
$\sigma_2 = \kappa \, \sigma_1$
consider the angular frequencies $\hat{\omega_1}$,
$\hat{\omega_2}$, $\hat{\omega_3}$ and $\hat{\omega_4}$ for which
these expressions assume their maxima over angular frequencies
according to Eqs.~(33) and~(38) in
Lindeberg \cite{Lin25-JCompNeurSci-orisel}
\begin{equation}
  \label{eq-omega1-spat}
    \hat{\omega}_{\varphi}  = \frac{1}{\sigma_1 \sqrt{\cos^2 \theta + \kappa^2 \sin^2 \theta}},
\end{equation}
\begin{equation}
  \label{eq-omega2-spat}  
  \hat{\omega}_{\varphi\varphi}
  = \frac{\sqrt{2}}{\sigma_1 \sqrt{\cos^2 \theta + \kappa^2 \sin^2 \theta}},
\end{equation}
and according to Eqs.~(40) and~(45) in the supplementary material of
Lindeberg \cite{Lin25-JCompNeurSci-spanelong}
\begin{equation}
  \label{eq-omega3-spat}
  \hat{\omega}_{\varphi\varphi\varphi}
  = \frac{\sqrt{3}}{\sigma_1 \sqrt{\cos^2 \theta + \kappa^2 \sin^2 \theta}},
\end{equation}
\begin{equation}
  \label{eq-omega4-spat}  
  \hat{\omega}_{\varphi\varphi\varphi\varphi}
  = \frac{2}{\sigma_1 \sqrt{\cos^2 \theta + \kappa^2 \sin^2 \theta}}.
\end{equation}
Given these preferred angular frequencies for the different orders of
spatial differentiation, let us next for the composed
integrated affine quasi quadrature
measures (\ref{eq-int-quasi-quad12-dir}),
(\ref{eq-int-quasi-quad1234-dir}) and
(\ref{eq-int-quasi-quad34-dir}) choose the geometric averages of the
respective components according to
\begin{equation}
  \label{eq-def-omega12}
  \hat{\omega}_{12} = \sqrt{\hat{\omega}_{\varphi} \, \hat{\omega}_{\varphi\varphi}},
\end{equation}
\begin{equation}
  \label{eq-def-omega1234}
  \hat{\omega}_{1234} =
  \sqrt{\hat{\omega}_{\varphi} \,
    \hat{\omega}_{\varphi\varphi} \,
    \hat{\omega}_{\varphi\varphi\varphi} \,
    \hat{\omega}_{\varphi\varphi\varphi\varphi}},
\end{equation}
\begin{equation}
  \label{eq-def-omega34}
  \hat{\omega}_{34} = \sqrt{\hat{\omega}_{\varphi\varphi\varphi} \,
    \hat{\omega}_{\varphi\varphi\varphi\varphi}},
\end{equation}
and adapt the angular frequency of the probing sine wave in
these ways to the respective quasi quadrature measures
${\cal Q}_{\varphi,12,\text{int}} L$ according to (\ref{eq-int-quasi-quad12-dir}),
${\cal Q}_{\varphi,1234,\text{int}} L $ according to
(\ref{eq-int-quasi-quad1234-dir}) and
${\cal Q}_{\varphi,34,\text{int}} L$ according to
(\ref{eq-int-quasi-quad34-dir}).
This adaptation of the angular frequency of the sine wave to
the internal parameters of the model of the complex cell
corresponds to probing the corresponding complex cells for
different values of the angular frequency $\omega$ and
then choosing the orientation selectivity curve for
the angular frequency $\hat{\omega}$ that leads to the maximum
response over all the frequencies $\omega$.

\begin{figure*}[hbtp]
  \begin{center}
    \begin{tabular}{c}
      \hspace {-17mm} ${\cal Q}_{\varphi,12,\text{int}} L$ \\
      \includegraphics[width=0.60\textwidth]{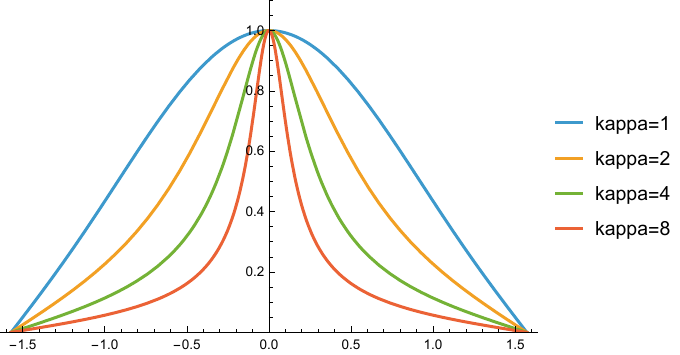} \\
      $\,$ \medskip \\
      \hspace {-17mm} ${\cal Q}_{\varphi,1234,\text{int}} L$ \\
      \includegraphics[width=0.60\textwidth]{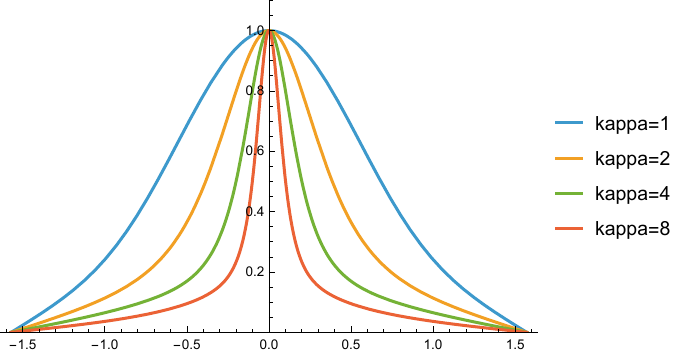} \\
      $\,$ \medskip \\
      \hspace {-17mm} ${\cal Q}_{\varphi,34,\text{int}} L$ \\
      \includegraphics[width=0.60\textwidth]{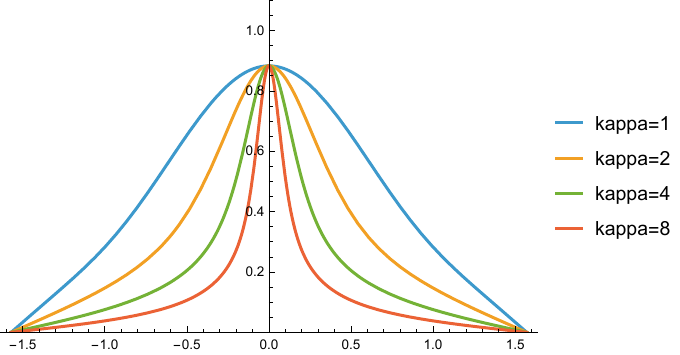} \\
     \end{tabular}
  \end{center}
  \caption{{\bf Graphs of the orientation selectivity curves for the generalized
    integrated affine quasi quadrature measures
    ${\cal Q}_{\varphi,12,\text{int}} L$ according to
    (\ref{eq-int-quasi-quad12-dir}),
    ${\cal Q}_{\varphi,1234,\text{int}} L$ according to
    (\ref{eq-int-quasi-quad1234-dir}) and
    ${\cal Q}_{\varphi,34,\text{int}} L$ according to
    (\ref{eq-int-quasi-quad34-dir})},
    and which combine integrated squared values of affine Gaussian
    derivative responses for different combinations of orders of
    integration $m \in \{ 1, 2, 3, 4 \}$ when applied to an ideal sine
    wave of the form (\ref{eq-sine-wave-model-spat-anal}) with
    the angular frequency of the sine wave adapted so as to evoke
    a maximally strong response over the angular frequencies according to 
    (\ref{eq-def-omega12}), (\ref{eq-def-omega1234}) and
    (\ref{eq-def-omega34}), respectively.
    The resulting orientation selectivity curves are shown for
    different values of the scale parameter ratio
    $\kappa = \sigma_2/\sigma_1 \in \{ 1, 2, 4, 8 \}$, which parameterizes the degree of
    elongation of the receptive fields.
    (Horizontal axes: orientation $\theta \in [-\pi/2, \pi/2]$.
     Vertical axes: Amplitude of the receptive field response relative
     to the maximum response obtained for $\theta = 0$.)}
  \label{fig-ori-sel-spat-anal}
\end{figure*}

Fig.~\ref{fig-ori-sel-spat-anal} shows the resulting orientation selectivity curves that
we then obtain for different values of the scale parameter ratio
$\kappa$, when using the relative integration scale
$\gamma = 1/\sqrt{2}$ for the spatial integration stage
and the weighting factor $C = 1/\sqrt{2}$ when combining
the responses for different orders $m$ of differentiation.
As can be seen from these graphs, for all the three generalized
integrated affine quasi quadrature measures, the orientation selectivity
curves become sharper with increasing values of the scale parameter
ratio $\kappa$. In this respect, these results are consistent with
the previously reported results in
Lindeberg
\cite{Lin25-JCompNeurSci-orisel,Lin25-JCompNeurSci-spanelong}.

Specifically, in relation to the previously recorded orientation
selectivity curves for biological neurons reported in
Nauhaus {\em et al.\/} \cite{NauBenCarRin09-Neuron},
and which span a variability in orientation selectivity from
wide to narrow orientation selectivity properties, these results
are consistent with what would be the case if the receptive fields in
the primary visual cortex would span a variability in the degree of
elongation, as proposed as a working hypothesis in
Lindeberg \cite{Lin23-FrontCompNeuroSci} Sec.~3.2.1
and further investigated in Lindeberg \cite{Lin25-JCompNeurSci-spanelong}.

\subsection*{Orientation selectivity histograms for the generalized
  idealized models of complex cells}

The previous analysis is {\em qualitative\/} in the sense that it shows that
the orientation selectivity curves become sharper for increasing
values of the scale parameter ratio $\kappa$, and also in the respect
that a variability in the orientation selectivity properties is
consistent with an underlying variability in the degree of elongation
of the receptive fields.

\begin{figure*}[hbtp]
  \begin{center}
    \begin{tabular}{c}
      {\em\small (a) Biological complex cells (n = 184)} \\
      \includegraphics[width=0.60\textwidth]{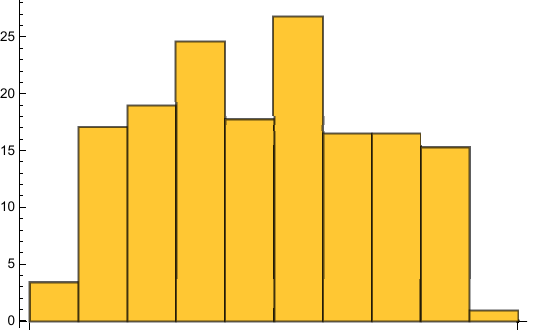}
      \\
      $\,$ \medskip \\
      {\em (b) Histogram over $|R|$ for }${\cal Q}_{\varphi,12,\text{pt}} L$ \\
      \includegraphics[width=0.60\textwidth]{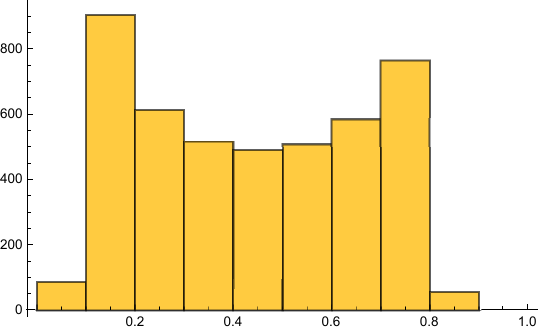}
    \end{tabular}
  \end{center}
  \caption{{\bf Orientation selectivity histogram of the resultant of
      biological complex cells accumulated by
      Goris {\em et al.\/}\ \protect\cite{GorSimMov15-Neuron} with comparison
      to histogram of the resultant from previous orientation
      selectivity analysis of maximally simplified models of complex cells
      in Lindeberg \protect\cite{Lin25-JCompNeurSci-spanelong}.}
  (Horizontal axes: 10 quantized bins over the resultant $|R| \in [0, 1]$.
  Vertical axes: bin counts.)}
  \label{fig-hist-resultant-Goris-Lindeberg}
\end{figure*}

To aim at a more {\em quantitative\/} analys, let us compare the result of our
idealized integrated models of complex cells with the quantitative
measurements of orientation selectivity histograms reported by
Goris {\em et al.\/}\ \cite{GorSimMov15-Neuron}.
They accumulated histograms of the absolute value of
the resultant $|R|$ with the underlying complex-valued resultant of
an orientation selectivity curve $r(\theta)$ of the form
\begin{equation}
  \label{eq-def-resultant}
  R = \frac{\int_{\theta = - \pi}^{\pi} r(\theta) \, e^{2 i \theta} d\theta}
                {\int_{\theta = - \pi}^{\pi} r(\theta) \, d\theta}.
\end{equation}
Fig.~\ref{fig-hist-resultant-Goris-Lindeberg}a gives a schematic
illustration of the results that they obtained, reflecting a
significant variability in wide {\em vs.\/}\ narrow orientation
selectivity properties for different biological complex cells.
Fig.~\ref{fig-hist-resultant-Goris-Lindeberg}b shows a corresponding
prediction of a histogram of the resultant of the orientation
selectivity curves obtained in
Lindeberg \cite{Lin25-JCompNeurSci-spanelong},
based on the pointwise quasi quadrature measure
${\cal Q}_{\varphi,12,\text{pt}} L$ in (\ref{eq-quasi-quad-dir}),
and assuming that the scale parameter ratio $\kappa$
would be uniformly distributed on a logarithmic scale over
the interval $[1/\kappa_{\max}, \kappa_{max}]$ for the
arbitrary choice of $\kappa_{max} = 8$.

As can be seen from the comparison between the biological results and
the idealized modelling results, the distribution over the resultant
$|R|$ is somewhat biased towards both smaller and larger values
of $|R|$, specifically regarding the peaks in the histogram at the
bins $|R| \in [0.1, 0.2]$ and $|R| \in [0.7, 0.8]$,
compared to the neurophysiologically accumulated resultant
histograms. A natural question to ask is hence if this behaviour would
be different if using more developed models of complex cells, that
also comprise a spatial integration stage and derivatives of higher
order than 2.

\begin{figure*}
  \begin{center}
    \begin{tabular}{c}
      {\em (a) Graph of resultant $|R(\kappa)|$ for} ${\cal  Q}_{\varphi,12,\text{int}} L$ \\
      \includegraphics[width=0.60\textwidth]{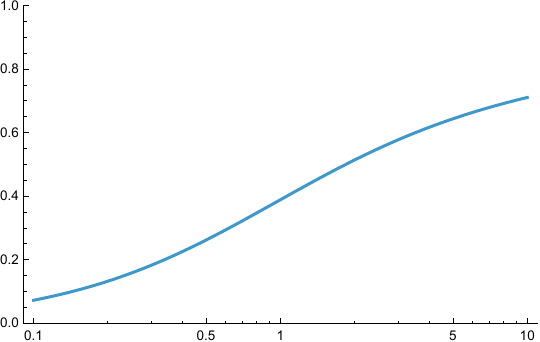}  \\
      $\,$ \medskip \\
      {\em (b) Graph of resultant $|R(\kappa)|$ for} ${\cal  Q}_{\varphi,1234,\text{int}} L$ \\
      \includegraphics[width=0.60\textwidth]{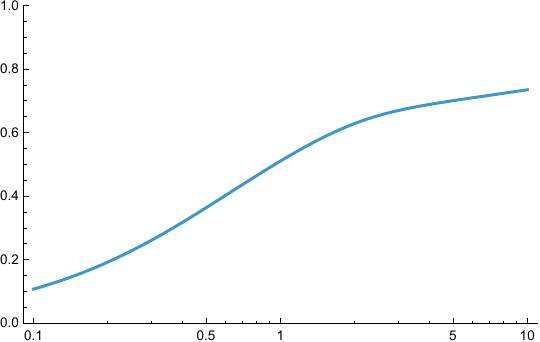} \\
      $\,$ \medskip \\
      {\em (c) Graph of resultant $|R(\kappa)|$ for} ${\cal Q}_{\varphi,34,\text{int}} L$ \\
      \includegraphics[width=0.60\textwidth]{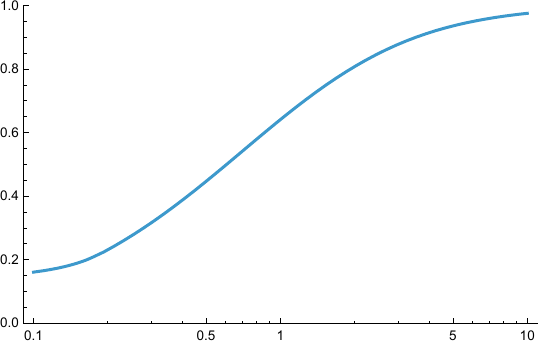} 
    \end{tabular}
  \end{center}
  \caption{{\bf Graphs of the resultant $|R(\kappa)|$ for the
      idealized models of complex cells
      ${\cal Q}_{\varphi,12,\text{int}} L$ according to (\ref{eq-int-quasi-quad12-dir}),
      ${\cal  Q}_{\varphi,1234,\text{int}} L$ according to (\ref{eq-int-quasi-quad1234-dir})
      and
      ${\cal Q}_{\varphi,34,\text{int}} L$ according to
      (\ref{eq-int-quasi-quad34-dir}).}
    (Horizontal axes in: scale parameter ratio $\kappa$.
    Vertical axes: resultant $|R|$.)} 
  \label{fig-graph-resultant}
\end{figure*}

Fig.~\ref{fig-graph-resultant} shows the
result of computing the resultant measure $|R|$ as a function
of the scale parameter ratio $\kappa = \sigma_2/\sigma_1$
for each one of the integrated affine quasi quadrature measures
${\cal Q}_{\varphi,12,\text{int}} L$,
${\cal Q}_{\varphi,1234,\text{int}} L$ and ${\cal Q}_{\varphi,34,\text{int}} L$
according to (\ref{eq-int-quasi-quad12-dir}),
(\ref{eq-int-quasi-quad1234-dir}) and
(\ref{eq-int-quasi-quad34-dir}).

\begin{figure*}
  \begin{center}
    \begin{tabular}{c}
      {\em (a) Histogram over $|R|$ for} ${\cal Q}_{\varphi,12,\text{int}}
      L$ \\
      \includegraphics[width=0.50\textwidth]{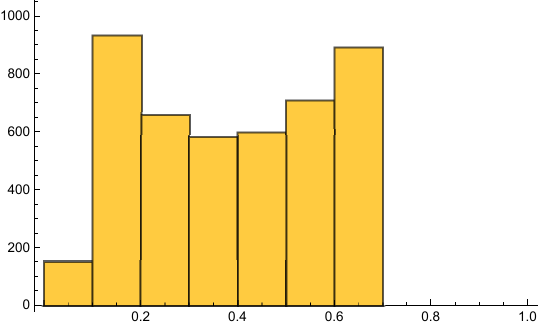}
      \\
      $\,$ \medskip \\
      {\em (b) Histogram over $|R|$ for} ${\cal
        Q}_{\varphi,1234,\text{int}} L$ \\
      \includegraphics[width=0.50\textwidth]{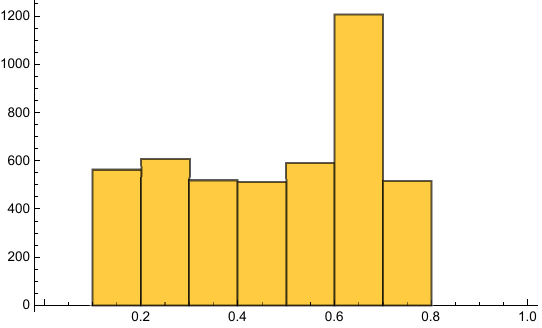}
      $\,$ \medskip \\      
      {\em (c) Histogram over $|R|$ for} ${\cal Q}_{\varphi,34,\text{int}} L$ \\
      \includegraphics[width=0.50\textwidth]{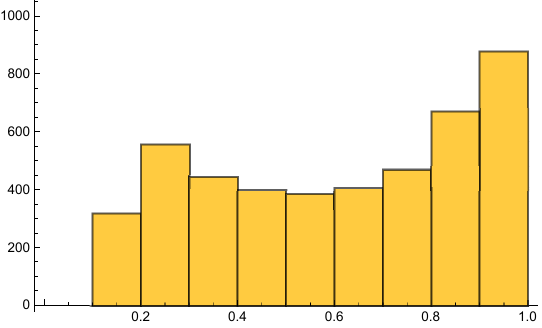}
      \\
      $\,$ \medskip \\           
      {\em (d) Combined histogram over $|R|$ for}
      ${\cal Q}_{\varphi,12,\text{int}} L$ {\em and} ${\cal Q}_{\varphi,34,\text{int}} L$\\
      \includegraphics[width=0.50\textwidth]{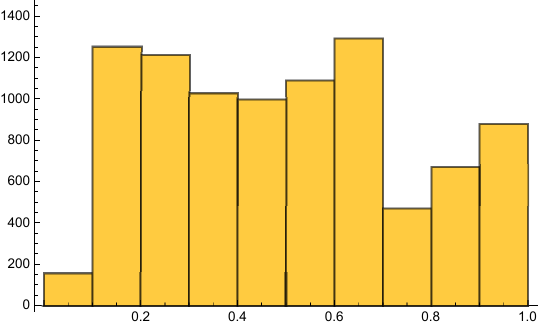} \\
    \end{tabular}
  \end{center}
  \caption{{\bf Orientation selectivity histograms for the 
   resultant $|R|$ for (a--c) the integrated
   affine quasi quadrature measures ${\cal Q}_{\varphi,12,\text{int}} L$,
   ${\cal Q}_{\varphi,1234,\text{int}} L$ and ${\cal Q}_{\varphi,34,\text{int}} L$
   according to (\ref{eq-int-quasi-quad12-dir}),
   (\ref{eq-int-quasi-quad1234-dir}) and
   (\ref{eq-int-quasi-quad34-dir}), when assuming a uniform
   distribution over a logarithmic parameterization of the scale
   parameter ratio $\kappa$, as well as (d)
   combined histogram of
   ${\cal Q}_{\varphi,12,\text{int}} L$ and ${\cal  Q}_{\varphi,34,\text{int}} L$,
   when assuming equal numbers of complex cells for these two types.}
   (Horizontal axes:
   10 quantized bins over the resultant $|R| \in [0, 1]$.
   Vertical axes: bin counts.)}
  \label{fig-hist-result-graphs}
\end{figure*}

Figs.~\ref{fig-hist-result-graphs}a--c show
corresponding resultant histograms for each of these
integrated affine quasi quadrature measures, when assuming a uniform
distribution over a logarithmic parameterization of the scale
parameter ratio $\kappa$ over the interval
$[1/\kappa_{\max}, \kappa_{\max}]$ for the again arbitrary
value of $\kappa_{\max} = 8$. Such a logarithmic distribution
constitutes a natural default prior for a strictly positive
variable according to Jaynes \cite{Jay68-SMC}.
Fig.~\ref{fig-hist-result-graphs}d shows
a combined histogram of the integrated affine quasi quadrature
measures ${\cal Q}_{\varphi,12,\text{int}} L$ and
${\cal Q}_{\varphi,34,\text{int}} L$, when assuming an equal
number of idealized complex cells for these two types.

As can be seen from these histograms, the use of the integrated
affine quasi quadrature measures ${\cal Q}_{\varphi,12,\text{int}} L$,
${\cal Q}_{\varphi,1234,\text{int}} L$ and ${\cal Q}_{\varphi,34,\text{int}} L$
leads to resultant histograms with different shapes than for
pointwise quasi quadrature measure ${\cal Q}_{\varphi,12,\text{pt}} L$
previously studied in
Lindeberg \cite{Lin25-JCompNeurSci-spanelong}.
Specifically, except for the peak at the bin corresponding
to $|R| \in [0.6, 0.7]$, the resultant histogram of
${\cal Q}_{\varphi,1234,\text{int}} L$ is more uniform
and without any bias the bias towards either smaller
values of the resultant $|R|$ towards the bins
$|R| \in [0.1, 0.2]$ and $|R| \in [0.7, 0.8]$
as for the resultant histogram of the
pointwise quasi quadrature measure ${\cal Q}_{\varphi,12,\text{pt}}
L$.

Additionally, when including third- and fourth-order terms,
the integrated affine quasi quadrature measures
${\cal Q}_{\varphi,1234,\text{int}} L$ and ${\cal Q}_{\varphi,34,\text{int}} L$
lead to accumulations to the bins
$|R| \in [0.8, 0.9]$ and $|R| \in [0.9, 1.0]$,
while those bins are not really reached by the
pointwise quasi quadrature measure ${\cal Q}_{\varphi,12,\text{pt}}
L$. In these respects, this analysis suggests that the mechanisms
of spatial integration and inclusion of higher-order terms than
mere first- and second-order terms may be important to
more quantitatively model the orientation selectivity properties
of complex cells. Furthermore, one may speculate if a
suitable reformulation of the non-linearities in the
composed integrated affine quasi quadrature measure
${\cal Q}_{\varphi,1234,\text{int}} L$ could move
the peak for the bin $|R| \in [0.6, 0.7]$
to the bin positions $|R| \in [0.3, 0.4]$
and $|R| \in [0.5, 0.6]$ that lead to peaks in
the resultant histogram for the biological complex cells.

Based on these results, we therefore propose
to (i)~include an explicit variability over the degree of elongation
of the receptive fields,
(ii)~include third- and fourth-order models of simple cells
in addition to previous use of first- and second-order models
of simple cells and (iii)~integrate non-linear transformations
of such receptive field responses over extended regions in
image space, when modelling the functional properties of complex
cells. Specifically, we propose that the items (i) and (iii) above should
be of much wider generality than restricted to computational models based
on affine quasi quadrature measures, and thereby also apply to other
families of computational models of complex cells.

\subsection*{Explicit predictions for further modelling of complex cells}
\label{sec-predictions}

More explicitly, based on the above results, we can thus state the
following general predictions:

\medskip

\noindent
{\bf Prediction~1 (Flexibility in elongation would lead to better
  approximation properties when modelling individual complex cells):}
Given a sufficiently large set of biological
complex cells, if the receptive fields of those are modelled by
computational models, then such models that involve a flexibility in
the degree of elongation of the underlying computational primitives
would lead to better approximation properties than for computational
models that do not involve a flexibility in the degree of elongation.

\medskip

\noindent
{\bf Prediction~2 (Variability in elongation over populations of
  complex cells):}
If the receptive fields sufficiently large set of complex cells are
modelled by computational models that involve a flexibility in the
degree of elongation of the underlying computational primitives, then
such model fitting would lead to a substantial variability in the
degree of elongation of the receptive fields over a sufficiently large population of
modelled biological complex cells.

\medskip

\noindent
{\bf Prediction~3 (Spatial integration as a computational mechanism in
  complex cells):}
Given a sufficiently large set of biological complex cells, if their
receptive fields are modelled by computational models, then models
that are based on spatial integration of underlying computational
primitives would lead to better approximation properties than
computational models that do not comprise any spatial integration.

\medskip

\noindent
Specifically, we propose that these predictions could be explored and
influence further modelling of complex cells in terms of
mathematically based image primitives.

\subsection*{Explicit suggestion to extension of methodology for
  experimentally probing the orientation selectivity of complex cells
  that may comprise a variability in the degree of elongation between
  different visual neurons.}
\label{sec-ext-meth-probe-ori-sel}

Furthermore, to allow for better distinctions between the validity of
different types of computational models for complex cells,
we propose to extend the experimental methodology for measuring
the orientation selectivity properties of biological neurons to
instead of (i)~choosing a single angular frequency for the probing
sine wave for generating the visual stimuli for each spatial image
orientation, instead (ii)~performing a simultaneous
{\em two-parameter variation\/} over
both the angular frequency and the image orientation.
In such a way, the experimental data ought to be better suited for
handling biological receptive fields with a variability in elongation.
The reason for this, is that the generated visual stimuli would then
better probe the dependency of two characteristic inherent spatial
scales of the biological receptive fields compared to using a single
inherent spatial scale for the visual stimuli.

As described in more detail in Lindeberg
\cite{Lin25-JCompNeurSci-orisel} Section~7, the choice of the angular
frequency for sine wave for probing the orientation selectivity
properties of visual neurons can significantly affect the shapes of
the resulting orientation selectivity curves, and should thus warrant
specific consideration.

\section*{Summary and discussion}
\label{sec-summ-disc}

We have presented a set of three new integrated affine quasi
quadrature measures to model the functional properties of complex
cells, and analyzed their orientation selectivity properties, based
on the assumption that the receptive field shapes should span a variability
over the degree of elongation of the simple cells that form the
input to these models of complex cells.

This analysis has been performed in three ways; in terms of:
(i)~orientation selectivity curves,
(ii)~graphs of the resultant $|R|$ as function of the scale parameter
ratio $\kappa = \sigma_2/\sigma_1$ between the scale parameter
$\sigma_2$ in the direction perpendicular to the preferred
orientation of the receptive field and the scale parameter $\sigma_1$
in the direction of the preferred orientation of the receptive field,
and (iii)~histograms of the resultant $|R|$ when assuming a
logarithmic distribution over the scale parameter ratio $\kappa$.

Specifically, by qualitative comparisons with the biological
results by Nauhaus {\em et al.\/}\ \cite{NauBenCarRin09-Neuron},
regarding a significant variability in orientation selectivity
properties of biological neurons from wide to narrow orientation
selectivity properties, and to
Goris {\em et al.\/}\ \cite{GorSimMov15-Neuron},
regarding orientation selectivity histograms over the resultant
$|R|$, we have found that these results are consistent with
a previously proposed hypothesis in
Lindeberg \cite{Lin23-FrontCompNeuroSci}
further investigated in
Lindeberg \cite{Lin25-JCompNeurSci-spanelong}
that the receptive fields in the primary visual cortex should span
a significant variability in the degree of elongation of the
receptive fields.
In this respect, the results are consistent with what becomes
a natural consequence of stating desirable properties of
an idealized vision system, that the receptive fields should
be covariant under the natural geometric image transformations.
In such a context, covariance properties of the family of receptive fields
enable more accurate estimates of cues to the 3-D structure
of the world, as the image data used as input to the vision
system undergo significant variabilities, as caused by
variations of the viewing conditions, such as the distance
and the viewing direction between objects in the world and
the observer.

Specifically, by simulating the orientation selectivity histograms that result
from the presented new integrated affine quasi quadrature measures
${\cal Q}_{\varphi,12,\text{int}} L$,
${\cal Q}_{\varphi,1234,\text{int}} L$ and ${\cal Q}_{\varphi,34,\text{int}} L$
according to (\ref{eq-int-quasi-quad12-dir}),
(\ref{eq-int-quasi-quad1234-dir}) and
(\ref{eq-int-quasi-quad34-dir}),
when combined with an assumption of a uniform distribution
over the logarithm of the scale parameter ratio $\kappa = \sigma_2/\sigma_1$
of the receptive fields, we have found that the extensions of
the previous pointwise quasi quadrature measure
${\cal Q}_{\varphi,12,\text{int}} L$ according to
(\ref{eq-quasi-quad-dir})
offer ways of changing the shapes of the predicted orientation
selectivity histograms regarding both how uniform the predicted
histograms will be in relation to the previously recorded
biological orientation selectivity histograms by
Goris {\em et al.\/}\ \cite{GorSimMov15-Neuron}
and regarding the span of values of the resultant $|R|$
they cover.

Thus, we propose to: (i)~include a variability over the degree of
elongation of the receptive fields when modelling the computational
function of complex cells, (ii)~include the mechanisms of spatial integration
and including receptive field responses of higher order than 2 when
modelling complex cells, and (iii)~use similar criteria to match
predicted orientation selectivity histograms to biological orientation
selectivity histograms, as used in the
presented analysis, to evaluate also other types of
computational models for complex cells.

Let us finally remark that it should most likely be the case that the non-linear behaviour of
complex cells may be more complex than the computational mechanisms
used in the proposed idealized models in terms of integrated affine
quasi quadrature measures. The overall purpose with this work is on
the other hand to demonstrate that the orientation selectivity
properties of the proposed idealized models of complex cells can be
analyzed with a structurally similar methodology as used for probing
the orientation selectivity properties of biological receptive fields.
From this viewpoint, the comparison to the biological orientation
selectivity histogram accumulated by
Goris {\em et al.\/}\ \cite{GorSimMov15-Neuron}
is specifically to show that the gross behaviour in terms of an underlying
distribution of receptive field shapes of different elongation can be
used to reflect gross properties of the biological measurements.
Our intention in this respect is to stimulate further tests
of more complex computational models of complex cells,
based on assuming distributions of the underlying receptive field
shapes over the degree of elongation.
The set of more explicit predictions in the section
``\nameref{sec-predictions}'' are specifically aimed at providing a
guide to such further research.

The proposed extension of the methodology for characterizing the
orientation selectivity of visual neurons in the section
``\nameref{sec-ext-meth-probe-ori-sel}'', by instead of performing a
one-parameter variation over the orientation of the probing sine wave
instead performing a two-parameter variation over both the angular
frequency and the orientation of the probing sine wave, is intended to
provide richer experimental data that could better distinguish between
the explanatory properties of different types of computational models
of complex cells.

From the viewpoint of quasi quadrature models to be used for
addressing computational tasks in computer vision, it should on
the other hand also be mentioned that a hierarchical network
constructed by applying a substantially simplified quasi quadrature model
of complex cells in cascade can lead to quite reasonable results on
computer vision benchmarks (Lindeberg \cite{Lin20-JMIV});
see also the closely related work by
Riesenhuber and Poggio \cite{RiePog99-Nature},
Serre {\em et al.\/} \cite{SerWolBilRiePog07-PAMI}
and Pant {\em et al.\/}
\cite{PanRodBenWarSer24-CognCompNeuroSci},
who construct hierarchical networks for computer vision tasks from sets of
idealized models of simple and complex cells coupled in cascade.
These studies in computational vision do thus demonstrate that
computational mechanisms structurally closely related to the
affine quasi quadrature measures proposed and studied in
this work can support spatial recognition tasks on visual data.


%
%
%


\end{document}